\documentclass[a4paper,11pt]{article}

\usepackage{jcappub}
\usepackage{graphicx}
\usepackage{amsmath}
\usepackage{amsfonts}
\usepackage{amssymb}
\usepackage{epsfig}
\newcommand{\be}{\begin{equation}}
\newcommand{\ee}{\end{equation}}

\newcommand{\ba}{\begin{eqnarray}}
\newcommand{\ea}{\end{eqnarray}}

\title{ Axions and high-energy cosmic rays:\\
Can the relic axion density be measured?}

\author[a,1]{
D. Espriu,%
\note{On leave 
of absence from DECM and ICCUB, Universitat de Barcelona}}

\affiliation[a]{CERN, 1211 Geneva, Switzerland,}
\author[b]{F. Mescia}
\author[b]{and A. Renau}
\affiliation[b]{Departament d'Estructura i Constituents de la
Mat\`eria and\\ Institut de Ci\`encies del Cosmos (ICCUB)\\
Universitat de
Barcelona,\\
Mart\'\i ~i Franqu\`es 1, 08028 Barcelona, Spain}

\abstract{
In a previous work we investigated the propagation of fast moving charged particles in a
spatially constant but slowly time dependent pseudoscalar background, such as the one
provided by cold relic axions. The background induces cosmic
rays to radiate in the low-energy spectrum. While the energy loss
caused by this mechanism on the primary cosmic rays is negligible, we
investigate the hypothetical detection of the photons radiated
and how they could provide an indirect way of
verifying the cosmological relevance of axions. Assuming that
the cosmic ray flux is of the form $J(E)\sim E^{-\gamma}$ we find that
the energy radiated follows a distribution
$k^{-\frac{\gamma-1}{2}}$ for proton primaries, identical to the Galaxy synchrotron
radiation that is the main background, and
$k^{-\frac{\gamma}{2}}$ for electron primaries, which in spite of this sharper decay
provide the dominant contribution in the low-energy spectrum.
We discuss possible ways to detect this small diffuse contribution thereby leading
to a potential direct detection of the cold axion background.
}

\arxivnumber{1010.2589}
\emailAdd{espriu@ecm.ub.es}
\emailAdd{mescia@ub.edu}
\emailAdd{arencer@gmail.com}

\begin{document}
%%%%%%%%%%%%%%%%%%%%%%%%%%%%%%%%%%%%%%%%%%%%%%%%%%%%%%%%%%%%%%%%%%%%%%%

\maketitle

\section{Introduction}

In this work we shall determine the flux of photons emitted by
high-energy cosmic rays travelling through an extremely diluted
pseudoscalar condensate oscillating in time. This
background could be provided by cold relic axions resulting
from vacuum misalignement in the early universe, which constitute
at present a viable possibility to explain the dark matter density
of the universe\cite{axions}. Detection of this radiation would constitute a
strong indirect evidence of the existence of the axion background.

Provided that the reheating temperature after inflation is below the
Peccei-Quinn transition scale\cite{PQ}, the axion collective
field evolves in later times as
\be
a(t)= a_0 \cos m_a t, \qquad {\bf k} =0
\ee
\be
\rho\simeq a_0^2 m_a^2
\ee
\be
\rho\simeq 10^{-30}{\rm g} {\rm cm}^{-3}\simeq 10^{-10} {\rm eV}^4,
\qquad
\rho_*\simeq 10^{-24}{\rm g} {\rm cm}^{-3}\simeq 10^{-4} {\rm eV}^4.
\ee
The last figure refers to the presumed axion density in galactic halos,
extending from 30 to 100 kpc\cite{density}.

The axion background thus
provides an extremely diffuse concentration of a pseudoscalar condensate
and one may be led to conclude that it is, except for its
gravitational effects, totally irrelevant.
However, the photon density associated to the microwave background
radiation is also very low and yet it has an impact on ultra-high energy
cosmic rays imposing the GZK cutoff\cite{GZK}. Consequently it seems natural
to investigate the effect of the axion background on the propagation
of highly energetic charged particles.

In \cite{aemr} it was seen that the time-varying axion background induces
a small amount of Lorentz violation that makes possible the
existence of 'cosmic ray Bremsstrahlung' processes, such as
$p\to p \gamma$ or $e \to e \gamma$
(forbidden in a Lorentz invariant theory) provided that the initial particle
of mass $m$ had an energy
\be
E > E_{th}= \frac{2m m_\gamma}{\eta}.
\ee
In the above expression $m_\gamma$ is a medium-induced effective photon mass.
Current bounds indicate that $m_\gamma < 10^{-18}$ eV\cite{photon}, but while we expect
the value of $m_\gamma$ not to be exactly zero it will likely be well
below this experimental bound.
In \cite{aemr} we used as reference value $m_\gamma \sim 10^{-18}$ eV. We shall
return to this issue below.

The quantity $\eta$ is the parameter characterizing Lorentz violation.
It appears in the following way.
The interaction of photons
with the axion background is described by the piece in the lagrangian
\be
\Delta{\cal L}= g_{a\gamma\gamma} \frac{\alpha}{2\pi} \frac{a}{f_a}\tilde F F
= -  g_{a\gamma\gamma} \frac{\alpha}{\pi} \frac{a_0 m_a}{f_a}
\sin (m_a t)\, \epsilon^{ijk} A_i F_{jk}
\ee
Popular models such as DFSZ\cite{DFSZ}  and KSVZ\cite{KSVZ}
all give $g_{a\gamma\gamma}\simeq 1$. Current observational bounds\cite{axions}
indicate that the coupling and preferred mass range are
\be
f_a > {\cal O }(10^{7}) {\rm ~GeV}, \qquad
10^{-1} {\rm eV} > m_a > 10^{-6} {\rm  ~eV},
\ee
although these bounds are based on a number of cosmological/astrophysical assumptions
and are somewhat weak\cite{masso}. Direct experimental bounds on the axion couplings only indicate
$f_a > 10$ TeV \cite{CHARM}. For
pseudo-Goldstone bosons related to the strong CP problem\cite{PQ} the approximate
relation $f_a m_a \simeq f_\pi m_\pi$ should hold\cite{gkr}.

If the momentum of a particle propagating in this background is large, ${\bf p}>> m_a$,
it makes sense to treat the axion background adiabatically and consider $\eta$
approximately constant:
\be
\Delta{\cal L}= \frac14 \eta \epsilon^{ijk} A_i F_{jk},
\ee
with $\eta_\alpha=(\eta,0,0,0)$  where the ``constant'' $\eta$ will change sign with a period $1/m_a$.
Numerically,
\be
\vert \eta\vert \simeq \alpha\frac{a_0 m_a}{f_a}= \alpha \frac{\sqrt{\rho_*}}{f_a}
\simeq  10^{-20}{\rm eV},
\ee
or less. For the
so called axion-like particles (see e.g. \cite{nt} for a recent proposal in connection
with dark matter) $f_a$ is actually unrelated to the mass and thus not bound by astrophysical
processes.

In the next section we quote the results on the kinematical limits concerning the
process $p\to p\gamma$, possible in a Lorentz non-invariant background, which will be needed in
the following discussion. In section 3 we compute the radiation probability and discuss the characteristics of
the comics ray flux needed to determine the intensity of the
radiation produced  by the axion-induced Bremsstrahlung mechanism proposed. In section 4 we
study the feasibility of the detection of the emitted radiation.

Our conclusions can be briefly summarized as follows. The dominant contribution to the radiation
yield via this mechanism comes from electron (and positron) cosmic rays.
If one assumes that the power spectrum of the cosmic
rays is characterized by an exponent $\gamma$ then the produced radiation has an spectrum
$k^{-\frac{\gamma-1}{2}}$ for proton primaries, which becomes $k^{-\frac{\gamma}{2}}$ for electron primaries.
The dependence on the key parameter $\eta\sim \frac{\sqrt\rho_*}{f_a}$
comes with the exponent $\eta^{\frac{1+\gamma}{2}}$ and $\eta^{\frac{2+\gamma}{2}}$ for protons and electrons, respectively.
However for the regions where the radiation yield is largest electrons amply dominate. We have assumed that
the flux of electron cosmic rays
is uniform throughout the Galaxy and thus identical to the one observed in our neighbourhood, but relaxing
this hypothesis could provide an enhancement of the effect by a relatively large factor.
The effect for the lowest wavelengths where the atmosphere is transparent and for values of $\eta$ corresponding
to the current experimental limit is of ${\cal O}(10^{-1})$ mJy. This is at the limit of sensitivity
of antenna arrays that are already currently being deployed and thus a possibility worth exploring.

\section{Summary of known results}

We shall review here some of the results obtained in \cite{aemr}. Consider the process
\be
p({\bf p}) \to p({\bf p}-{\bf k})\gamma({\bf k})
\ee
Let us first consider the case $m_\gamma=0$.
(see \cite{aemr} for details). We denote $k=\vert {\bf k}\vert$ and
assume that $\eta >0$ in what follows. The process is possible only for one
polarization of the final photon, which gets reversed if $\eta < 0$ so there
is no loss of generality in assuming a specific sign for $\eta$.
For the threshold energy of the cosmic ray and the kinematical limits on the radiated
photon we have
\be
E_{th}=0
\ee
\be
k_{min} =\eta, \quad {\rm ~for } \cos\theta=-\eta/2p
\ee
\be
k_{max}= \frac{E^2}{p+\frac{m_p^2}{\eta}}, \quad {\rm  ~for } \cos\theta=1
\ee
Note that $k_{max}\simeq E$ for $E\gg m_p^2/\eta$ and that $k_{max}\simeq \eta E^2/m_p^2$ for
$E\ll m_p^2/\eta$.

Let us next consider the case $m_\gamma >0$
\be
E_{th}\simeq \frac{2m_\gamma m_p}{\eta}
\ee
\be
k(\theta_{max})\simeq \frac{2m_\gamma^2}{\eta}(1 - 3 \frac{p
m_\gamma^2}{E^2\eta})\stackrel{E>>E_{th}}{\longrightarrow} \frac{2m_\gamma^2}{\eta},
\qquad
\sin^2\theta_{max} \to \frac{\eta^2}{4m_\gamma^2}
\ee
$\theta_{max}$ can be very small; photons are emitted in a narrow cone if $\eta< m_\gamma$ and
more isotropically otherwise.

In the opposite extreme, for zero angle, there are two solutions
\be
k_{+}(0)\simeq \frac{E^2\eta + pm_\gamma^2 + E\sqrt{E^2\eta^2-4m_p^2m_\gamma^2+2p\eta m_\gamma^2}}
{2p\eta + 2 m_p^2}
\stackrel{E>>E_{th}}{\longrightarrow} \frac{E^2}{p+\frac{m_p^2}{\eta}}
\ee
which is the same result obtained before, and
\be
k_{-}(0)\simeq \frac{E^2\eta + pm_\gamma^2 - E\sqrt{E^2\eta^2-4m_p^2m_\gamma^2+2p\eta m_\gamma^2}}
{2p\eta + 2 m_p^2}
\stackrel{E>>E_{th}}{\longrightarrow}\frac{m_\gamma^2}{\eta}
\ee
\be
k_-(0)<k(\theta_{max})<k_+(0) \nonumber
\ee

The rate of energy loss of the cosmic ray was also computed in \cite{aemr}
\be
\frac{dE}{dx} = - \frac{1}{v}\, \int d\Gamma(Q) w(Q)
\ee
\be
\frac{dE}{dx}= - \frac \alpha2\frac1{p^2}\int\,kdk [-\frac12(m_\gamma^2+\eta k)+p^2(1-\cos^2\theta)]
\ee
There are two relevant limits
\be
E \ll \frac{m_p^2}{\eta}\longrightarrow\frac{dE}{dx}=-\frac{\alpha\eta^2 E^2}{4m^2_p}.
\ee
\be
E\gg \frac{m_p^2}{\eta}\longrightarrow\frac{dE}{dx}=-\frac{\alpha\eta}{3}E
\ee

Notice that there are two key scales in this problem:
$ E_{th}\simeq 2m_\gamma m_p/\eta$ and $m_p^2/\eta $, the cross-over energy, where
$dE/dx$ changes behaviour; clearly $m_p^2/\eta \gg E_{th}$.
For energies $E \gg m_p^2/\eta$
\be
E(x)= E(0) \exp{- \frac{\alpha\eta}{3}x},
\ee
giving a mean free path
of ${\cal O}(1)$ pc. The `axion shield' would indeed be very effective at such
enormous energies. However due to the smallness of $\eta$ the crossover 
scale $m_p^2/\eta$ is many orders of
magnitude larger than the highest energy rays measured and the
above restriction on the mean free-path is not relevant. 
Even cosmic rays just below the GZK cut-off
of $10^{20}$ eV are
well below the cross-over scale $m_p^2/\eta$. In this regime, which is the relevant one, 
the expression for $E(x)$ is
\be
E(x)= \frac{E(0)}{1+\frac{\alpha\eta^2}{4m_p^2} E(0) x},
\ee
giving a much weaker suppression. From this expression and the fact that we detect (likely)
extragalactic rays of large energy we can set
at present the largely irrelevant bound
\be
\eta < 10^{-14}{\rm ~eV}
\ee
(recall that we expect $\eta\sim 10^{-20}$ eV or less).
It is peculiar to see that for extremely large distances $E(x)\sim \frac{1}{x}$ independently of their
primary energy but this regime is never reached even at the largest cosmic scales that are observable, so it remains
a curiosity. The net effect of the oscillating pseudoscalar background on cosmic ray propagation is
truly negligible.

\section{Radiation yield}

Let us turn to the radioemission
due to the axion-induced Bremsstrahlung.

For primary protons, using $m_\gamma= 10^{-18}$ eV and $\eta= 10^{-20}$ eV as indicative
values and the usual GZK cut-off, there would be electromagnetic activity in the region of the spectrum
\be
10^{-16} {\rm ~eV} (0.024 {\rm ~Hz}, \lambda= 1.2\times  10^7 {\rm ~km}   ) <  k  <
100 {\rm ~eV} (24 {\rm ~PHz}, \lambda= 12 {\rm ~nm}).
\ee
Before jumping prematurely to conclusions we have to estimate the energy yield which
will in fact be quite small at high energies.

For primary electrons, which are much rarer in number but radiate more (see the expressions
in the previous section), there would be
activity in the range
\be
10^{-16} {\rm ~eV}  < k  < 400 {\rm ~MeV},
\ee
assuming again $m_\gamma= 10^{-18}$ eV and $\eta= 10^{-20}$ eV and a cut-off similar
to the one of  protons. This last point is very questionable since the spectrum of electrons reaching
the Earth seems to bend down
around $\sim 10$  TeV\cite{electrons}; however the issue is still unclear.
As we will see the intensities at such high frequencies are very low anyway so the uncertainties
about the high-energy part of the electron cosmic ray spectrum are of little consequence.
Note that $m_\gamma$ affects only the lower limit of the above ranges and that
the kinematical limits on $k$ are proportional to $\eta$.
We shall eventually set $m_\gamma=0$.

In order to compute the radiation yield we shall need to estimate the number of cosmic rays
and their differential emission rate into photons of wave vector $k$, $d\Gamma/dk$. This latter
quantity was determined in \cite{aemr}
\be
\frac{d\Gamma}{dk}=\frac\alpha8\frac1{k\omega}\left[A(k)+B(k)E^{-1}+C(k)E^{-2}\right]
\theta(\frac{E^2\eta}{m^2}-k),
\ee
with
\be
A(k)=4(\eta k - m_\gamma^2),\quad
B(k)=4\omega(m_\gamma^2-\eta k),\quad
C(k)=-2m_\gamma^2k^2+2\eta k^3-m^4_\gamma-\eta^2k^2+2m_\gamma^2\eta k,
\ee
where
\be
\omega=\sqrt{m_\gamma^2-\eta k+k^2},
\ee
and $m$ is the mass of the charged particle.
Although given for an arbitrary value of $m_\gamma$ for completeness, it makes
sense to set $m_\gamma=0$ in the decay rate. The error is insignificant;
$m_\gamma$ is only relevant in the kinematics.

Let us consider a surface element $dS_0$ in space and consider the number of photons
radiated with wave vector $k$ by the cosmic rays crossing that surface element
within a time interval $dt_0$ and having an energy between $E$ and $E+dE$.
The number of such cosmic rays (protons or electrons) per unit surface will be
\be
d^3 N= J(E) dE dS_0 dt_0.
\ee
$J(E)$ is the usual cosmic ray flux; there is one for each type of cosmics.
The $d^3N$  cosmic rays will eventually radiate at a time $t$, unrelated
to $t_0$, and they will yield a number of photons with a given wave vector $k$
given by the usual differential decay formula
\be
d^5 N_\gamma = d^3 N \frac{d\Gamma(E,k)}{dk} dk dt= J(E) \frac{d\Gamma(E,k)}{dk} dE dk dt_0 dS dt.
\ee
$N_\gamma$ is dimensionless. $J(E)$ is expressed in units of eV$^{-1}$ m$^{-2}$ s$^{-1}$.

Now, assuming uniformity and isotropy of the cosmic rays we can safely assume that
the flux is the same for any such surface element $dS$ (indeed we have already set
$dS_0=d S$ in the above expression), and for any time interval $dt_0$
and integrate over $t_0$ obtaining a factor $t(E)$ equivalent to the average
lifetime of a cosmic ray of energy $E$. Therefore
\be
\frac{d^3 N_\gamma}{dk dS dt}=\int_{E_{th}}^\infty dE \; t(E) J(E)\frac{d\Gamma(E,k)}{dk}\qquad
E_{th}=2\frac{m_{p,e}m_\gamma}{\eta}. \label{yield}
\ee
Note that the units of $d^3 N_\gamma / dk dS dt$ are the same as those of $J(E)$.

Observations indicate that cosmic rays exhibit an energy spectrum of the form
\be
J(E)= N_i E^{-\gamma_i}
\ee
with $\gamma_i \simeq 3$. For protons we shall use the parametrization given in\cite{rays}. All energy units
in what follows are given in electronvolts.
\be
J_p(E)=\left\{\begin{array}{ll}
                     5.87 \cdot 10^{19} E^{-2.68} &  10^9\le E\le4\cdot10^{15} \\
                     6.57 \cdot 10^{28}E^{-3.26} & 4\cdot10^{15}\le E\le4\cdot10^{18} \\
                     2.23  \cdot 10^{16}E^{-2.59} & 4\cdot10^{18}\le E\le2.9\cdot10^{19} \\
                     4.22  \cdot 10^{49}E^{-4.3} &  E\ge2.9\cdot10^{19}
                   \end{array}
\right.
\label{cosmic1}
\ee
For electrons (less well measured, but typically 1\% of the proton flux)\cite{electrons}
\be
J_e(E)=\left\{\begin{array}{cc}
                     5.87 \cdot 10^{17} E^{-2.68} & E\le5\cdot10^{10} \\
                     4.16 \cdot 10^{21} E^{-3.04} & E\ge5\cdot10^{10}
                   \end{array}
\right.
\label{cosmic2}
\ee
Units are eV$^{-1}$ m$^{-2}$ s$^{-1}$ sr$^{-1}$ as stated.
We shall consider in what follows that $ E_{th} > 10^9$ eV  as the flux of cosmics below that
energy is likely to be influenced by local effects of the solar system. The previous parameterizations
describe the flux of protons at all measured energies with good precision and roughly describes
the one of leptons, which is more poorly known. The form of the electron flux turns out quite
relevant so our ignorance about the lepton flux is quite regrettable as it has
a substantial impact in our estimation of the radiation yield.

Note that the above ones are values measured locally in the inner solar system. It is known that
the intensity of cosmic rays increases with distance from the sun because the
modulation due to the solar wind makes more difficult for them to reach us, particularly so for electrons.
Therefore the above values have to be considered as lower bounds for the flux
which may be up to $\sim 10$ times larger in the nearby interstellar medium. In addition, the hypothesis of homogeneity
and isotropy hold for proton cosmic rays, but not necessarily for electron
cosmic rays. Indeed because cosmic rays are deflected by magnetic fields they follow a nearly random trajectory within
the Galaxy. Collisions of cosmic rays having large atomic number with the interstellar medium
sometimes produce lighter unstable radioactive isotopes. By measuring their abundance we know that on average a hadronic
cosmic ray spends about 10 million years in the galaxy before escaping into intergalactic space. This ensures
the uniformity of the flux, at least for protons of galactic origin. On the contrary, electron cosmic rays
travel for approximately 1 kpc on average before being slowed down
and trapped. However, because $l\sim \sqrt{D(E)t}$ ($D(E)$ is
the diffusion coefficient of the random walk) 1 kpc corresponds to a typical age
of a electron cosmic ray  $\sim 10^5$ yr\cite{kin}, a lot less than protons.
In addition, the lifetime of an electron cosmic ray depends
on the energy in the following way
\be
t(E)\simeq 5 \times 10^5 (\frac{1 \ {\rm TeV}}{E})\, {\rm yr}=\frac{T_0}E \label{ageee},
\ee
with $T_0 \simeq 2.4 \times 10^{40}$ if $E$ is measured in eV.
To complicate matters further, it has been argued that the local interstellar flux of electrons is not
even representative of the Galaxy one and may reflect the electron debris from a nearby supernova
$\sim 10^4$ years ago\cite{age}.

To get an estimate we will replace in the integral $t(E)\to T$ and assume the value $T=10^7$ yr for
protons and use (\ref{ageee}) for electrons.
The measured photon energy flux $I(k)$ per unit wave vector, measured per
unit surface per unit time and per sr will then be
\ba
I(k)&=&\omega(k)\int_{E_{min}(k)>E_{th}}^\infty dE\ t(E) J(E)\frac{d\Gamma}{dk}\cr
&=&\frac{\alpha}{8k}\int_{E_{min}(k)>{th}}^\infty dE\ t(E) N_i
\left[ A(k)E^{-\gamma_i}
+B(k)E^{-(\gamma_i+1)}
+C(k)E^{-(\gamma_i+2)}\right]\cr
&=&\frac{\alpha}8\frac{T}{k}\sum_i\left[
A(k)\frac{E^{-\gamma_i+1}}{-\gamma_i+1}
+B(k)\frac{E^{-\gamma_i}}{-\gamma_i}
+C(k)\frac{E^{-\gamma_i-1}}{-\gamma_i-1}\right]_{E_i^{initial}}^{E_i^{final}},
\ea
%\be
%I(k)=\omega(k)\int_{E_{min}(k)>E_{th}}^\infty dE\ t(E) J(E)\frac{d\Gamma}{dk}
%\nonumber
%\ee
%\be
%=\frac{\alpha}{8k}\int_{E_{min}(k)>{th}}^\infty dE\ t(E) N_i
%\left[ A(k)E^{-\gamma_i}
%+B(k)E^{-(\gamma_i+1)}
%+C(k)E^{-(\gamma_i+2)}\right]\nonumber
%\ee
%\be
%=\frac{\alpha}8\frac{T}{k}\sum_i\left[
%A(k)\frac{E^{-\gamma_i+1}}{-\gamma_i+1}
%+B(k)\frac{E^{-\gamma_i}}{-\gamma_i}
%+C(k)\frac{E^{-\gamma_i-1}}{-\gamma_i-1}\right]_{E_i^{initial}}^{E_i^{final}},
%\ee
with $E_{min}(k)=\sqrt{\frac{m^2 k}{\eta}}$.
In the previous expression the labels 'initial' and 'final' refer to the
successive energy ranges where the different parameters $N_i$ and $\gamma_i$ are
applicable. The above result applies to protons; for electrons the spectrum is
is reduced by one additional power of the energy.

Numerically, it is straightforward to see that
 the whole contribution is
dominated by the initial point $E_{min}(k)$. Furthermore only the term proportional
to $A(k)=4\eta k$ in the decay rate is numerically relevant. Then
\be
I_\gamma^p(k)\simeq \frac{\alpha \eta T}{2} \frac{J_p(E_{min}(k)) E_{min}(k)}{\gamma_{min}-1}.\label{approx}
\ee
and
\be
I_\gamma^e(k)\simeq \frac{\alpha \eta T_0}{2} \frac{J_e(E_{min}(k))}{\gamma_{min}}\label{approx1}.
\ee
Energies are all expressed in eV.
The value $\gamma_{min}$ is to be read from (\ref{cosmic1}) and (\ref{cosmic2}) depending
on the range where $E_{min}(k)$ falls.
The above approximate formula
(\ref{approx}) and (\ref{approx1}) reproduces the exact result within an accuracy that is sufficient for our purposes.

\section{Results and discussion}

We should now settle the discussion on the value of $m_\gamma$. The best observational
limits on the effective photon mass come from  measurements of the
Jovean magnetic field of magneto-hydro dynamics of the solar wind. They are obviously
measurements at very long wavelengths. Even more stringent (but not accepted as a direct
limit by the Particle Data Group) is a 10$^{-27}$ eV bound derived from the existence
of the galactic magnetic field.

Theoretically we expect that the dominant contribution to the effective photon mass is induced by
the electron density that
is expected to be at most of the order of $n_e=10^{-7}$ cm$^{-3}$. Photons would pick up a mass
\be
m_\gamma^2\simeq  4\pi\alpha\frac{n_e}{m_e}.
\ee
This expression gives $m_\gamma= 10^{-15}$ eV. However a first consideration is that
the density of free electrons is of course not uniform, but significant only around some active
regions with larger plasma densities. More importantly, because the density of free
electrons is so low, it takes
photons of very low momentum to `see'
a collective effect due to the density of electrons. Typically the distance
for the collective effect of the
electron plasma to induce an effective mass
will have to be $>> n_e^{-\frac13}$. Since we will typically be interested in photons
with a shorter wavelength
it seems safe to conclude that $m_\gamma$ has to be set to its fundamental value, namely zero.

If the above considerations hold the value assumed for $E_{th}$ comes not from
kinematical considerations but from
the practical need to ensure
that all cosmic rays included in the determination of the radiation yield due to the
axion-induced Bremsstrahlung have traveled a large distance and thus have had enough time
to contribute to the electromagnetic yield.
Cosmic rays from the solar system normally reach a maximum energy of 1 GeV, and very rarely
10 GeV\cite{solar}. We therefore take $E_{th}=1$ GeV both for electrons and protons. In this
way we can set, if the detection of the effect is positive, a reliable lower bound on $\eta$.
Since $E_{th} < E_{min}(k) = \sqrt{\frac{m^2 k}{\eta}}$ we are sure that
photons with $k > 10^{-7}$ eV were radiated off cosmic rays {\em not} of solar origin.
We take $k=10^{-7}$ eV as the reference scale as this is approximately
the minimum wave vector at which the atmosphere is transparent to electromagnetic
radiation, even though the signal is higher for lower frequency photons. This corresponds to
30 MHz, a band in which an extensive antenna array (LWA) is already being commisioned\cite{lwa}.
In the same range of extremely low frequencies the Square Kilometer Array (SKA) project 
could cover a the range from 70 to 10,000 MHz with enormous sensitivity (see below)
\cite{SKA}.

As a result of the previous considerations we expect the following measured intensities (flux
densities)
from the axion-induced Bremsstrahlung. First of all, the dominant contribution comes
from electrons
\be
I_\gamma^e(k)\simeq
3\times 10^2 \times \left(\frac{\eta}{10^{-20}~{\rm eV}}\right)^{2.52}
  \left(\frac{k}{10^{-7}\ {\rm eV}}\right)^{-1.52}
{\rm ~m}^{-2}\, {\rm s}^{-1}\, {\rm sr}^{-1} .\label{approxyield}
\ee
For protons
\be
I_\gamma^p(k)\simeq
6  \times \left( \frac{T}{10^7 \ {\rm yr}}\right) \left(\frac{\eta}{10^{-20}~{\rm eV}}\right)^{1.84}
 \left(\frac{k}{10^{-7}\ {\rm eV}}\right)^{-0.84}
{\rm ~m}^{-2}\, {\rm s}^{-1}\, {\rm sr}^{-1}.\label{approxyield1}
\ee
In a way it is unfortunate that the dominant contribution comes from electron cosmic rays
because they are still poorly understood.
Note that $I(k)$ has the dimensions of energy per unit wave vector per unit surface
per unit time. In radioastronomy the intensity, or energy flux density, is commonly
measured in Jansky
(1 Jy = 10$^{-26}$ W Hz$^{-1}$ m$^{-2}$ sr$^{-1}$ $\simeq 1.5\times 10^7$ eV eV$^{-1}$  m$^{-2}$ s$^{-1}$ 
sr$^{-1}$).

The expected overall
intensity is shown in Figure 1 in a doubly logarithmic scale for a very wide
range of wave vectors (many of them undetectable) and for the
reference values for $T,\eta$ indicated in (\ref{approxyield}) and (\ref{approxyield1}).
It should be emphasized that this is really only
a rough estimate of the background radiation provided by  cosmic rays of galactic origin due
to axion-induced Bremstrahlung. We have
assumed very conservatively that the flux measured in the inner solar system is a good representative of
the average abundance of cosmic rays in the galaxy, but this is almost certainly an
underestimate due to our peripheral position in the galaxy and the relatively short reach of
electron cosmic rays.

\begin{figure}
\centering
\epsfxsize=400pt
\epsfbox{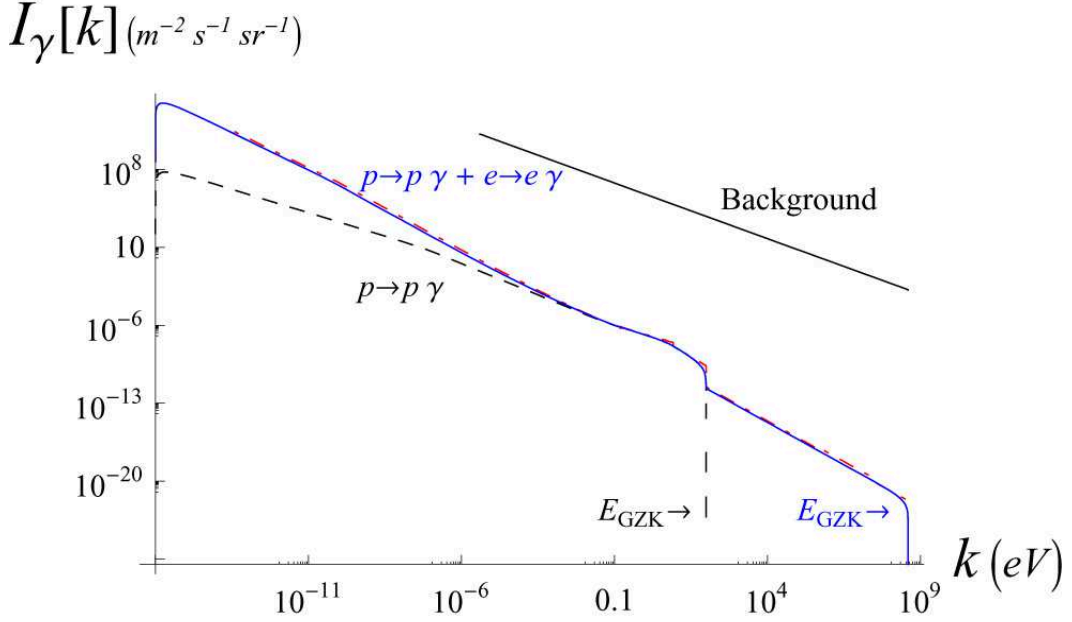}
%\includegraphics[
%natheight=1.3716in, natwidth=1.516in, height=1.4062in, width=1.5523in]
%{Cosmic_new.pdf}
\caption{Total intensity $I_\gamma= I_\gamma^p + I_\gamma^e$ expected to be measured as a consequence of the axion Bremsstrahlung
effect discussed here. Units are in m$^{-2}$ s$^{-1}$ sr$^{-1}$. The total yield is the
external envolvent and it is dominated by
electrons for a wide range of frequencies. The figure is plotted using the exact formulae (solid line). The
proton contribution is shown separately (dashed line). The approximate expressions discussed in the text
are shown in dotted-dashed line (nearly invisible). For comparison the approximate galactic radio background 
(basically from electron synchrotron radiation) is shown\cite{kraus}. Note that the radio background is 
not well measured at present below
10 MHz but there are indications suggesting a marked decrease below 3 MHz. In the 100 MHz region the axion induced signal
is about nine orders of magnitude smaller than the background.}
\end{figure}

In order to see whether this flux is measurable from the Earth or not one has to determine
the diffuse noise perceived by the receiver in the appropriate wavelength, known identified
background sources, and of course
take into account the atmosphere transparency at that radiation wavelength. As it is well
known\cite{opacity}, the atmosphere is transparent to radiation in the terrestrial
microwave window ranging from approximately 6 mm (50 GHz,  $2
\times10^{-4}$ eV) to 20 m (15 MHz, $6 \times10^{-8}$ eV), becoming
opaque at some water vapor and oxygen bands and less transparent as
frequency increases up to 1 THz. The current technology allows for radio
detection from space up to 2 THz (e.g. with the Herschel Space
Observatory\cite{Graauw}) but the low receiver sensitivity at
frequencies in the submillimeter band ($>300$ GHz) could be an issue.
There are further considerable narrower windows in
the near infrared region from 1 $\mu$m (300 THz, 1.2 eV) to around 10 $\mu$m (30 THz, 0.12 eV).
This region can be explored by space missions.
The atmosphere blocks out completely the emission in the UV and X-Ray region
corresponding to $\lambda < 600$ nm ($k > 80$ eV), a region that is actively being
explored by spaceborne missions.

If  $\lambda > 2.5$ m (0.8 GHz, $3 \times 10^{-6}$eV),
the galactic synchroton radiation noise increases rapidly difficulting
the detection of any possible signal. Note however that while
the power spectrum of the axion-related radiation from proton primaries is the same as
the one from the synchrotron radiation they produce\cite{synchro}, the
bulk of the Galaxy synchroton radiation is due to electrons whose
spectral power law describing the axion-induced Bremsstrahlung is different. In addition
there would be a difference between the galactic and the axion based
synchrotron emission anyway. In fact\footnote{We thank P. Planesas for pointing out this
possibility to us}, in areas of high galactic
latitude, where no local features superpose the broad galactic emission,
the measured spectral index is  $\sim - 0.5$ \cite{rr}.
Instead, the axion induced effect has a power $\sim - 1.5$ if we assume $\gamma\sim 3$.

The maximum observed values\cite{db} for the intensities are: 10$^4$ m$^{-2}$ s$^{-1}$ sr$^{-1}$ in
the X-Ray region and up to  from 10$^{10}$ to 10$^{14}$ m$^{-2}$ s$^{-1}$ sr$^{-1}$
in the radio, IR and UV regions
but the sensitivity of antenna arrays at very low frequencies such as the LWA\cite{lwa} can be as low
as 0.1 mJy $\simeq 10^3$  m$^{-2}$ s$^{-1}$ sr$^{-1}$ or even less.
Of particular interest for our purposes is the sensitivy that can be reached in the SKA antenna. This
can be estimated\cite{condon} assuming an integration time of 50 hrs at the lowest frequency
to be 650 nJy. This is clearly several orders below the expected size of the effect, even
assuming the worst possible case for the electron flux. Therefore, while the effect is below the
sensitivity of existing antennas it will be within the reach of several projects in construction
or under consideration\footnote{It may be worth noticing that the long standing 
project of setting up an antenna on the far side 
of the Moon\cite{moon} could reach sensistivities of $10^{-5}$ Jy or 
less, also providing enough sensitivity even for
pessimistic values of the electron flux. Such an antenna would of 
course not be limited by the atmosphere opacity
being sensitive -in principle- to even longer wavelengths.}.

Once it is clear that antennas can measure fluxes twelve orders below the dominant Galaxy synchrotron
radiation in the galactic plane, it is obvious that 
sensitivity to the axion signal (`only' nine orders below the average galactic noise) is not an issue, 
the real difficulty is to disentangle the effect from the background or foreground. For this
purpose the rather different power dependence should prove essential. The difference in power spectrum 
between the expected signal and the background is
even more marked for regions of high galactic latitude as already mentioned. Good angular
resolution will be essential too as observers looking for this signal will probably be interested
in focusing their instruments in region with low magnetic fields\footnote{Note that the
Galaxy magnetic field varies by about three orders of magnitude from $\mu$G to mG}, where 
synchrotron radiation
will be at a minimum  gaining several orders of magnitude in the signal-to-noise
ratio\footnote{The synchrotron radiation depends quadratically on the
magnetic field, hence a change of two orders of magnitude in the magnetic 
field represent a variation of four orders
in the amount of the synchrotron ratiation background}.

While it is obviously beyond the scope of this paper (and the expertise of the authors)
to present a definite proposal to measure the tiny
axion-induced Bremsstrahlung predicted in this work, we do conclude that it 
is conceivably within the reach of
a new generation of instruments specifically designed for exploration
of the long wavelength region. We do not exclude that it can be found in the exploration
of close extragalactic sources either. In both cases the main unknown is a detailed understanding
of the nature and spectrum of electron cosmic rays, an issue worth investigating by itself for a variety of
raisons.

Other comments pertinent here are the following. Firstly one should note that the effect discussed here
is a collective one. This is at variance with the GZK effect alluded in the introduction
- the CMB radiation is not a coherent one
over large scales. For instance, no similar effect exists for hot axions.
A second observation is that some of the scales that play a role in the present discussion
are somewhat non-intuitive (for instance the 'cross-over' scale $m_p^2/\eta$ or the
threshold scale $m_\gamma m_p/\eta$). This is due to
the non Lorentz-invariant nature of this effect. Also, it may look surprising at first that
an effect that has such a low probability may give a small but not ridiculously small contribution.
The reason why this happens
is that the number of cosmic rays is huge. It is known that they contribute to the energy density
of the Galaxy by an amount similar to the Galaxy's magnetic field\cite{adr}. Finally we would 
like to comment that the calculations presented here in the limit where the oscillations
are assumed to be adiabatic can be proven to be exact\cite{er}.

We hope that the present mechanism
can shed some light on the presumed relevance of cold axions as a dark matter candidate.

\section*{Acknowledgements}

We acknowledge the financial support from projects FPA2007-66665,
2009SGR502, Consolider CPAN CSD2007-00042 and FLAVIANET. We thank A de Rujula for
some comments on the manuscript concerning cosmic rays and, particularly, J.M. Paredes and P. Planesas
for comments and suggestions concerning the Galaxy synchrotron radiation.


\begin{thebibliography}{99}


\bibitem{axions}L.Abbott and P. Sikivie, Phys. Lett. 120B, 133 (1983);
M. Kuster, G. Raffelt and B. Beltran (eds), Axions: Theory, Cosmology
and Experimental Searches, Lecture Notes
in Physics 741 (2008).

\bibitem{PQ}R.D. Peccei and H.R. Quinn, Phys. Rev. Lett. 38, 1440 (1977)

\bibitem{density}E. W. Kolb and M. S. Turner, The Early Universe (Westview Press, 1990); Y. Sofue and V. Rubin,
Ann. Rev. Astron. Astrophys. 39, 137 (2001); S. J. Asztalos et al., Ap. Jour. 571, L27 (2002);
E.I. Gates, G. Gyruk and M.S. Turner, Ap. Jour. 449, L123 (1995)

\bibitem{GZK}G.T. Zatsepin, V.A. Kuz'min, Zh. Eks. Teor. Fiz., Pis'ma
Red.4 (1966)144; K. Greisen, Phys. Rev. Lett. 16, 748 (1966).

\bibitem{aemr} A. Andrianov, D. Espriu, F. Mescia and A. Renau, Phys. Lett. B684, 101 (2010). See
also A. Andrianov. D. Espriu, P. Giacconi and R. Soldati, JHEP 0909:057 (2009).

\bibitem{photon}C.Amsler et al. [Particle Data Group], Phys. Lett. B 667 (2008) 1.

\bibitem{DFSZ}M. Dine, W. Fischler and M. Srednicki, Phys. Lett. B, 104, 199 (1981); A.R. Zhitnitsky,
Sov. J. Nucl. Phys. 31, 260 (1980).

\bibitem{KSVZ}J. E. Kim, Phys. Rev. Lett. 43, 103 (1979); M. A. Shifman, A. I. Vainshtein and V. I. Zakharov,
Nucl. Phys. B 166, 493 (1980).

\bibitem{masso} J. Jaeckel, E. Masso, J. Redondo, A. Ringwald and F. Takahashi, Phys. Rev. D 75, 013004 (2007).

\bibitem{CHARM} F. Bergsma et al. [CHARM Collaboration], Phys. Lett. B 157, 458 (1985).

\bibitem{gkr} H. Georgi, D. Kaplan and L. Randall, Phys. Lett. B169, 73 (1986).

\bibitem{nt} Y. Nomura and J. Thaler, Phys. Rev. D79:075008, 2009.

\bibitem{electrons}K. Nakamura et al. (Particle Data Group), J. Phys. G 37, 075021 (2010).
M. Aguilar-Benitez et al [AMS-I Collaboration], Phys. Rep. 366, 331 (2002); M. Boezio et al. [CAPRICE Collaboration],
Astrophys. J. 532, 653 (2000). See also the
recent measurement by M. Ackermann et al [The FERMI LAT Collaboration], arXiv: 1008.3999v1,
confirming previous estimates.

\bibitem{rays}J. Abraham et al. [The Pierre Auger Collaboration], Phys. Lett. B 685, 239 (2010)
[arXiv:1002.1975 [astro-ph.HE]]; M. Amenomori et al [The Tibet AS$\gamma$ Collaboration], Astrophys. J.
678, 1165 (2008).

\bibitem{kin} N. Kawanaka, K. Ioka and M. Nojiri, Astrophys. J. 710 (2010) 958.

\bibitem{age} A. Casadei and V. Bindi, in Proceedings of the 28th International Cosmic Ray Conference,
Universal Academy Press, Tokyo, 2003.

\bibitem{solar} T.K. Gaisser, Cosmic Rays and Prticle Physics, Cambridge University Press (Cambridge, UK) 1990;
J.A. Simpson, Annual reviews of Nuclear and Particle Science, 33, 323 (1983).

\bibitem{lwa} S.W. Ellingson et al., Proceedings of the IEEE 97, 1421 (2009).
See also arXiv:1005.4332v1, to appear in IEEE Transactions on Antennas and Propagation.

\bibitem{SKA} R. Taylor, IAU Proc. 248 (2007). See also http://www.skatelescope.org/.


\bibitem{kraus} J.D. Kraus, Radio Astronomy, 2nd edition (1986), Cygnus-Quasar Books (New York, USA).


\bibitem{opacity} H.J. Leibe, Radio Science 20, 1069 (1985); see also
J. D. Krauss, Radio Astronomy, 2nd edition, Cygnus-Quasar (Powell, Ohio, USA) 1986.

\bibitem{Graauw} Th. de Graauw et al, Astronomy and Astrophysics 518, L6 (2010).


\bibitem{synchro} M.S. Longair, High Energy Astrophysics, vol. 2, Cambridge University Press, Cambridge (UK), 1994;
B.F. Burke and F. Graham-Smith, "An introduction to radio astronomy",
Cambridge University Press (Cambridge, UK), 1997;
See also E. Orlando, Gamma rays from interactions of cosmic-ray electrons, Ph. D. Thesis (2008),
http://www.imprs-astro.mpg.de/Alumni/Orlando-Elena.pdf.

\bibitem{rr} P. Reich and W. Reich, Astronomy and
Astrophysics 196, 211 (1988), Astronomy and Astrophysics Supplement 74, 7 (1988).

\bibitem{db} E. Dwek and M.K. Barker, Astrophys. J. 575, 7 (2002);
F.E. Marshall et al., Astrophys. J. 235, 4 (1980); E. Churazov et
al., Astron. Astrophys. 467, 529 (2007); F. Frontera et al., Astrophys. J. 666, 86 (2007).

\bibitem{condon} J. Condon, SKA memo 114, http://www.skatelescope.org/.

\bibitem{moon} P.Y. Bely et al, ``Very low frequency array on the lunar far side'', ESA report SCI(97)2, 
European Space Agency. 
 
%\bibitem{arcade}A. Kogut et al [The ARCADE 2 collaboration] arXiv: 0901.555v1, 0901.0562v1.

\bibitem{adr}See e.g.: A. Dar and A. De Rujula, Phys.Rept. 466,179 (2008).

\bibitem{er} D. Espriu and A. Renau, in preparation.


\end{thebibliography}
\end{document}